\documentclass[a4paper,10pt]{article}
\usepackage[english]{babel}
\usepackage{color}
\usepackage[margin=2cm,tmargin=2cm]{geometry}
\usepackage[nottoc]{tocbibind}
\usepackage{amsmath}
\usepackage{graphicx}
\usepackage{breqn}
\usepackage{color}
\usepackage{matht   ools}
\usepackage{multicol}
\usepackage{url}
\usepackage{fancyhdr}
\usepackage{csquotes}
\usepackage{wrapfig}
\usepackage[dvipsnames]{xcolor}
\usepackage{lipsum}
\usepackage{atbegshi}
\graphicspath{{Images/}{../Images}}
\usepackage{subfiles}
\usepackage[toc,page]{appendix}
 \usepackage{subcaption}
\usepackage{blindtext}

\usepackage[acronym]{glossaries}
\makeglossaries
\usepackage{acronym}
\usepackage{titling}
\usepackage{lipsum}

\pretitle{\begin{flushleft}\LARGE\bfseries}
\posttitle{\par\end{flushleft}}
\preauthor{\begin{flushleft}\Large}
\postauthor{\end{flushleft}}
\predate{\begin{flushleft}}
\postdate{\end{flushleft}}

\title{The Mu3e Experiment: Status and Short-Term Plans \large Contribution to the 25th International Workshop on Neutrinos from Accelerators}

\author{\normalsize\textbf{R.M. Amarinei\textsuperscript{1} for the Mu3e Collaboration}, \\ \vspace{0.3cm}
~\textsuperscript{1}Department of Nuclear and Particle Physics, University of Geneva, 
24 rue du Général-Dufour, Geneva, CH\\ 
~~~Email: robert.mihai.amarinei@cern.ch 
}

\begin{document}
\maketitle
\vspace{-0.7cm}
\hrule
\vspace{1cm}
\begin{abstract}
Mu3e is an experiment currently under construction at the Paul Scherrer Institute in Switzerland, designed to search for the Lepton Flavor Violating (LFV) decay \(\mu^+ \rightarrow e^+e^-e^+\). In extensions of the Standard Model (SM) that account for neutrino masses, this decay is theoretically allowed but occurs only through extremely rare loop processes, with a predicted branching ratio of approximately \(\mathcal{O}(10^{-54})\). Such a small probability implies that any observation of this decay would provide clear evidence for physics beyond the SM.  

The Mu3e experiment aims to probe the \(\mu^+ \rightarrow e^+e^-e^+\) decay with a sensitivity of approximately \(\mathcal{O}(10^{-15})\) in its Phase-1 and plans to achieve a sensitivity of  \(\mathcal{O}(10^{-16})\) after future upgrades. To reach its Phase-1 ambitious goals, Mu3e is going to use the most intense continuous muon beam in the world, generating 10$^{8}$ muon stops per second in the target placed at the center of the Mu3e. 

Mu3e will use three main technologies for particle detection. The tracking will done through ultra-thin (50 - 70 $\mu m$) pixel detectors based on MuPix11 sensors. These are high-voltage monolithic active pixel sensors (HV-MAPS) with a $\sim$ 23~$\mu$m spatial resolution. The timing will be done through scintillating fibres ($\sim$ 250 ps) and tiles ($\sim$ 40 ps), coupled to silicon photomultipliers and read out by MuTRiG3 ASICs. 
A triggerless DAQ system based on FPGAs will collect data from the detectors, which will then undergo reconstruction in a GPU filter farm. The assembly of the detectors has started, with a detector commissioning beam time planned for 2025. This document reports on the status of the construction, installation, and data-taking plans for the near future.
\end{abstract}

\newpage
\section{Physics motivation}

The decay \( \mu^+ \to e^+e^-e^+ \) (muon decaying into an electron and two positrons) is highly interesting in physics because it represents a lepton flavor-violating (LFV) process, which is forbidden in the Standard Model (SM) of particle physics under the assumption of massless neutrinos. Even with the discovery of neutrino oscillations (implying nonzero neutrino masses), the predicted branching ratio for such decays in the SM is exceedingly small, on the order of \( \mathcal{O}(10^{-54}) \). This suppression arises due to the tiny neutrino masses and the GIM (Glashow–Iliopoulos–Maiani) mechanism~\cite{Calibbi:2017uvl}. The observation of \( \mu^+ \to e^+e^-e^+ \) at any measurable rate would therefore be a clear signal of physics beyond the Standard Model (BSM). 

Along with the muon conversion to an electron (\( \mu^- N \to e^- N \)) and the muon decay into an electron and a photon (\( \mu \to e\gamma \)), the muon decay into two electrons and a positron (\( \mu \to eee \)) is one of the three so-called ``golden channels'' for studying rare muon decays. Each of these LFV decays is equally powerful as a probe for physics beyond the standard model, with different sensitivities to various operators and couplings in BSM physics. For example, \( \mu \to eee \) probes four-fermion contact interactions, whereas \( \mu \to e\gamma \) is more sensitive to dipole interactions mediated by new particles (e.g., SUSY particles or heavy gauge bosons).

Observing multiple LFV channels would allow physicists to disentangle the underlying new physics. This motivates a big experimental physics scene, with experiments optimized to study each golden channel: MEG and MEG II~\cite{MEG,MEGII:2018kmf} for the  \( \mu \to e\gamma \), COMET and Mu2e for $\mu^- N \rightarrow e^- N$~\cite{COMET,Mu2e}, and Mu3e~\cite{Mu3e:Cite} designed specifically for the muon decay to two positrons and an electron.

\section{The Mu3e Experiment}

Mu3e is an experiment currently being built at the Paul Scherrer Institute (PSI) in Switzerland, home to the most intense continuous muon beam in the world ($\pi E5$). This facility allows Mu3e to operate with the high muon rates required to achieve its ambitious sensitivity goals. The experiment is planned in two phases, with increasing sensitivity as upgrades are implemented. Phase 1 aims to reach a branching ratio sensitivity of \( \sim 10^{-15} \), using $\pi E5$, the existing high-intensity muon beam at PSI, which delivers \( 10^8 \) muons per second with a 28~MeV/c momentum. In Phase 2, Mu3e will extend its sensitivity down to \( 10^{-16} \), using the upgraded High-Intensity Muon Beamline (HiMB) which will deliver $10^9$  muons stops per second, with a similar momentum as the muons from Phase 1.

The layout of the Mu3e detector, together with the muon beam direction, is shown schematically in Fig.~\ref{fig:Mu3eGeneralScheme}. The muon beam enters the detector on a side and is stopped in a hollow double-cone target made of Mylar. This design is optimized for the efficient stopping of muons while minimizing material interactions that could lead to unwanted backgrounds. It also allows decay products (electrons and positrons) to exit the target with minimal scattering, ensuring precise tracking and momentum reconstruction. The Mu3e detector system is designed to achieve high spatial, temporal, and momentum resolution, relying on cutting-edge technologies.

\subsection{The Pixel Tracker}

\begin{figure}
    \centering
    \includegraphics[width=0.7\linewidth]{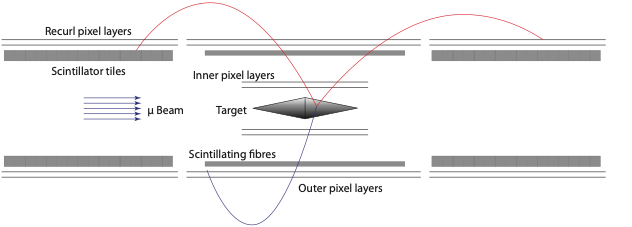}
    \caption{A detailed view of the Mu3e detector layers, with the muon beam directed toward the target positioned at the Mu3e core~\cite{Mu3e:Cite}.}
    \label{fig:Mu3eGeneralScheme}
\end{figure}

The pixel tracker uses High-Voltage Monolithic Active Pixel Sensors (HV-MAPS~\cite{HVMAPS}) to track the trajectories of electrons and positrons with high spatial precision (\( \sim 23~\mu \text{m} \)). This tracker is composed of multiple cylindrical detector layers surrounding the target (see Fig.~\ref{fig:Mu3eGeneralScheme}), providing detailed track reconstruction and vertex identification. The detector is divided into three main stations:  

\begin{itemize}  
    \item \textbf{Vertex Detector:} Surrounds the hollow muon stopping target, closest to the decay vertex.  
    \item \textbf{Central Outer Tracker:} Positioned further outward to enhance tracking and momentum measurements.  
    \item \textbf{Recurl Stations:} Located upstream and downstream of the central detector to detect particles recurling in the solenoidal magnetic field.  
\end{itemize}  

For the vertex detector, the sensors are built on \( 50~\mu\text{m} \)-thick silicon wafers, while the outer detector sensors use \( 70~\mu\text{m} \)-thick silicon wafers.  The vertex detector is constructed from single chips, with each chip having an active area of around  $20 \times 20~\text{mm}^2$. Including the readout area, the total chip size is approximately \( 23~\text{mm} \). Each pixel layer is composed of ladders, and each ladder in the vertex detector is built using six chips.  

The pixel detector achieves a position resolution of \( 23~\mu\text{m} \), ensuring precise tracking of decay products. Its efficiency exceeds \( 99\% \), and although not explicitly designed for timing, it provides a time resolution better than \( 20~\text{ns} \).  

The pixel detector dissipates approximately \( 215~\text{mW/cm}^2 \) and requires active cooling to manage the heat load. To minimize the material budget, the detector employs helium gas cooling, which introduces significantly less material compared to conventional cooling methods. For comparison, \( 1~\text{m} \) of air corresponds to approximately \( 0.33\%~X_0 \) of radiation length, while \( 1~\text{m} \) of helium corresponds to just \( 0.018\%~X_0 \)~\cite{RUDZKI2023168405}. This innovative approach ensures minimal disruption to the trajectories of decay products, maintaining the high precision of the detector.

\subsection{Timing Detectors }

In addition to the pixel tracker, Mu3e incorporates Scintillating Fibers (SciFi) and Scintillating Tiles (SciTile) for precise timing measurements. At the core of each detector is the MuTRiG, a silicon photomultiplier readout ASIC with high timing precision and high event rate capability~\cite{MuTRiG_cite}.

The SciFi, positioned in a very dense area just outside the vertex layers, provides timing information with a resolution of \( \sim 250~\text{ps} \). The building block of the SciFi detector is a so-called ribbon.   Each ribbon is constructed from 3 layers of \( 250~\mu\text{m} \) staggered scintillating fibers.  The entire SciFi detector comprises 12 ribbons, arranged to provide full \( 4\pi \) coverage. A single ribbon has a total thickness of \( 720~\mu\text{m} \), corresponding to a material budget of  $0.2\%~X_0$.  The SciFi detector achieves a timing resolution of approximately \( 250~\text{ps} \) maintaining an efficiency higher than 97\%~\cite{SciFi2017},\cite{SciFi2023},\cite{Demets}.   

To manage the heat dissipation and mitigate radiation damage to the photon sensors, the SciFi detector is cooled using a liquid cooling system. This  uses silicon oil at a temperature of $-20^\circ \text{C}$, which circulates through the support structure of the SciFi, thus effectively cooling the needed areas. 

The SciTile is the last detector in the particle path, located below the pixel recurl stations.  As its name suggests, the SciTile is made of (Ej-228 plastic) scintillator tiles of 6 x 6 x 5 mm$^3$, each one equipped with Hamamatsu Silicon Photomultipliers (SiPMs). Due to lower material budget constraints, the SciTi performs with a fantastic time resolution of $\sim 40~\text{ps}$ and an efficiency of 99\%.

\subsection{Backgrounds}

A significant challenge for Mu3e is the suppression of backgrounds, particularly combinatorial backgrounds arising from high beam rates. For instance, this includes scenarios such as two Michel decays where one positron undergoes Bhabha scattering within the detector. Another type of background comes from radiative decays, with processes like $\mu^+ \to e^+ e^- e^+ \nu_e \nu_\mu$ or
$ \mu \to e\nu\nu\gamma $, where the photon converts into an \( e^+e^- \) pair. Such processes are significant because if the neutrinos carry small energies, the decay mimics the Mu3e signal.  

Each subsystem of Mu3e is optimized to remove these backgrounds. The pixel tracker provides excellent spatial resolution to reconstruct the decay vertex, distinguishing true \( \mu \to eee \) decays from accidental combinations. Its momentum resolution ($ \leq 0.5\%$ targeted) enables kinematic consistency checks, rejecting events that do not match the expected decay topology. The scintillating fibers and tiles offer precise timing resolution, allowing discrimination of signal events from accidental pileup in the high-rate environment. Time-of-flight measurements help confirm particle origins and reduce backgrounds from out-of-time events. Finally, a solenoidal magnetic field bends charged particle trajectories, enabling precise momentum measurements to identify signal particles and reject high-momentum backgrounds. Mu3e employs a triggerless readout system, ensuring no events are lost and enabling sophisticated offline analysis for optimal background rejection.

\section{Status}

The Mu3e experiment has reached a significant milestone, with the detector design finalized and all subsystems either completed and ready for installation or in the final stages of construction. The installation process is currently underway, with the goals of taking data with cosmic rays (vertex + SciFi + SciTi) in early 2025 and achieving beam time later in 2025 for full detector commissioning. Later on, the first physics data-taking is estimated for 2026.

The vertex detector has undergone successful mechanical integration tests using a mock-up, validating the installation procedures. A fully operational two-layer vertex detector is being prepared and is expected to be deployed in early 2025. Approximately 50\% of the SciFi detector is ready for installation and commissioning. The remaining modules are currently being constructed and are expected to be delivered in early 2025. The SciTile detector group has already produced and tested several modules, with one available for the Spring 2025 cosmic run.  

The helium cooling system for the pixel detector has been constructed and successfully commissioned, ensuring effective thermal management with minimal material contribution. The liquid cooling system shared between the SciFi and SciTile detectors, has been constructed, and its full commissioning is scheduled for the upcoming months. The low-voltage and high-voltage connectors have been fully integrated into the detector cage, ensuring the readiness of Mu3e for operation. In addition, the wiring of the data acquisition system is currently in progress, providing the infrastructure required to handle the high data rates expected during operation.  

With the detector nearing completion and installation well underway, the Mu3e experiment is on track for its upcoming commissioning phases and early operational runs, marking a major step forward in the search for rare muon decays.

\bibliographystyle{unsrt}
\bibliography{references.bib}
\end{document}